\documentclass{emulateapj}

\usepackage{lscape}
\usepackage{rotate}
\usepackage{longtable}
\usepackage{natbib}
\shorttitle{Chandra Observations of Cl0023+0423 at $z\sim0.84$}
\shortauthors{Kocevski et al.}

\begin{document}

 \title{No Evidence of Quasar-Mode Feedback in a Four-Way Group Merger at $z\sim0.84$}
 \author{Dale D. Kocevski, Lori M. Lubin, Brian C. Lemaux, Roy R.
  Gal\altaffilmark{1}, Christopher D. Fassnacht, Gordon K. Squires\altaffilmark{2}}

\affil{Department of Physics, University of California, Davis, 1 Shields Avenue,  Davis, CA 95616}
\altaffiltext{1}{Institute for Astronomy, University of Hawaii, 2680 Woodlawn Dr., Honolulu, HI 96822}
\altaffiltext{2}{\emph{Spitzer} Science Center, M/S 220-6, California 
                 Institute of Technology, 1200 East California Blvd, Pasadena, CA 91125}
\email{kocevski@physics.ucdavis.edu}

\begin{abstract}

\end{abstract}

\keywords{galaxies: active --- galaxies: clusters: general --- X-rays: galaxies: clusters}

\begin{abstract}

We report on the results of a \emph{Chandra} search for evidence of triggered
nuclear activity within the Cl0023+0423 four-way group merger at
$z\sim0.84$.  The system consists of four interacting galaxy groups in the
early stages of hierarchical cluster formation and as such, provides a
unique look at the level of processing and evolution already
underway in the group environment prior to cluster assembly.  We present the
number counts of X-ray point sources detected in a field covering the entire
Cl0023 structure, as well as a cross-correlation of these sources with our
extensive spectroscopic database.   Both the redshift distribution and
cumulative number counts of X-ray sources reveal
little evidence to suggest the system contains X-ray luminous AGN in excess
to what is observed in the field population.   If preprocessing is underway
in the Cl0023 system, our observations suggest that powerful nuclear
activity is not the predominant mechanism quenching star formation and
driving the evolution of Cl0023 galaxies.  We speculate that this is due
to a lack of sufficiently massive nuclear black holes required to power such
activity, as previous observations have found a high late-type fraction among the 
Cl0023 population.  It may be that disruptive AGN-driven outflows only
become an important factor in the preprocessing of galaxy populations during
a later stage in the evolution of such groups and structures, when
sufficiently massive galaxies (and central black holes) have built up, but
prior to hydrodynamical processes stripping them of their gas reservoirs.

\end{abstract}

\section{Introduction}

There is now substantial evidence that environments of intermediate density,
such as galaxy groups, play an important role in the transformation of field
galaxies into the passively evolving populations found in galaxy clusters.
Several studies have found that group populations already exhibit reduced
star formation rates (SFR; Lewis et al.~2002; Gomez et al.~2003) and high
early-type fractions similar to those observed in denser environments
(Zabludoff \& Mulchaey 1998; Jeltema et al.~2007).  While the physical
mechanisms responsible for the preprocessing of galaxies in the group regime
are still heavily debated, several recent studies have reported an
overdensity of X-ray luminous Active Galactic Nuclei (AGN) on the outskirts
of clusters and within the substructure surrounding unrelaxed systems (D'Elia
et al.~2004; Cappelluti et al. 2005; Kocevski et al.~2009a,2009b; Gilmour et
al.~2009). These observations suggest that increased nuclear activity may be
triggered in such environments and that AGN-driven outflows may play a role
in suppressing star formation within galaxies during cluster assembly.
Indeed the increased dynamical friction within groups and their low relative
velocity dispersions make them conducive to galaxy interactions which can
trigger such activity (Hickson 1997; Canalizo \& Stockton 2001) and recent
hydrodynamical simulations suggest that merger-triggered AGN feedback can
have a profound effect on the gas content and star formation activity of
their host galaxies (Hopkins et al.~2007; Somerville et al.~2008). 

Since the mass density of virialized structures increases with redshift,
mergers are expected to play an even greater role in the group environment in
the past.  Therefore, if galaxy interactions and subsequent AGN feedback are
driving a significant portion of the preprocessing found in intermediate
density environments, we may expect to find an overdensity of AGN in high
redshift groups in the early stages of hierarchical cluster formation.  In
this Letter we report on \emph{Chandra} observations of one such system, the
Cl0023+0423 (hereafter Cl0023) four-way group merger at $z=0.84$ (Lubin et
al.~2009a).  The Cl0023 structure consists of four interacting galaxy groups
which simulations suggest are the direct progenitors of a future massive
cluster.  As such, the system provides a unique look at the level
of processing and evolution already underway in the group environment prior
to cluster assembly. 

To search for evidence of triggered nuclear activity within the Cl0023
structure, we present the number counts of X-ray point sources detected in a
field covering the entire system, as well as a cross-correlation of these
sources with our extensive spectroscopic database.  Surprisingly, we find
no evidence for an overdensity of X-ray detected point sources in the
direction of the Cl0023 groups.  We discuss the implications of this finding
on the role of AGN feedback in regulating galaxy evolution in such
structures. We also examine possible explanations for the lack of increased
nuclear activity in the system.  Throughout this Letter we assume a
$\Lambda$CDM cosmology with $\Omega_{m} = 0.3$, $\Omega_{\Lambda} = 0.7$, and
$H_{0} = 70$ $h_{70}$ km s$^{-1}$ Mpc$^{-1}$.

\begin{figure*}
\epsscale{1.1}
\plotone{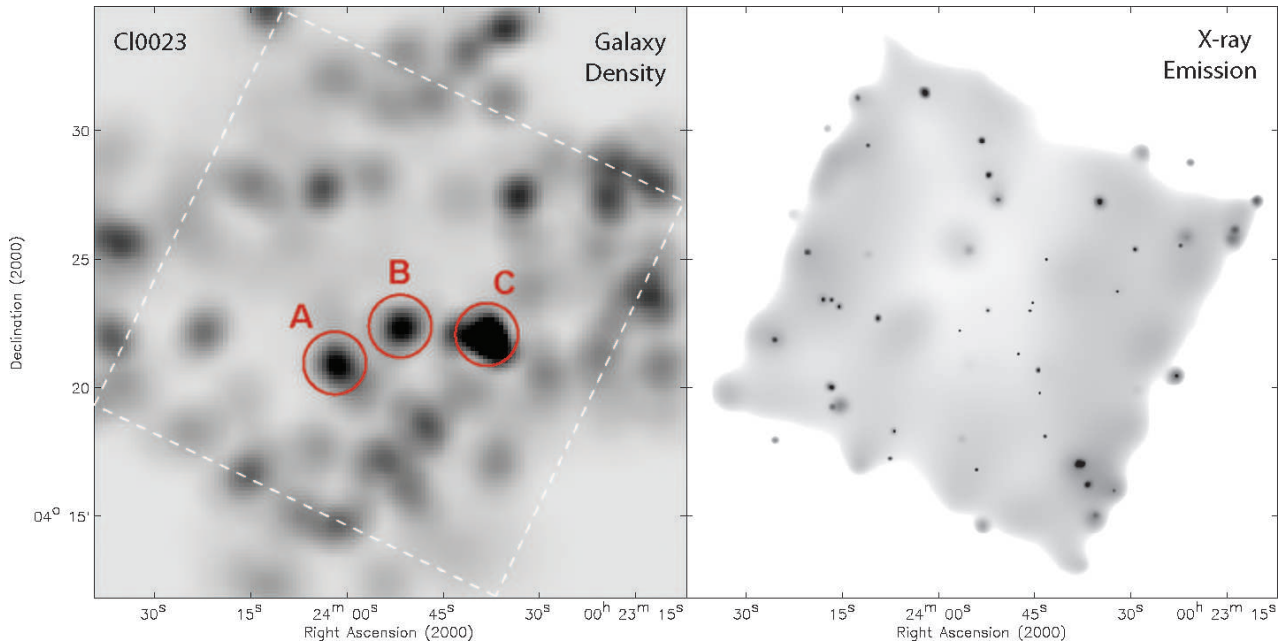}
\caption{(\emph{left}) Adaptively smoothed density map of
  color-selected red galaxies in the Cl0023 field.  Three density peaks that
  correspond with four spectroscopically confirmed galaxy groups are marked.
  Adapted from Lubin et al.~(2009a).  (\emph{right}) Adaptively smoothed, ACIS-I image of the Cl0023 field
  in the soft X-ray band (0.5-2 keV). \label{fig-densmap}}
\end{figure*}

\vspace{0.5in}
\section{The Cl0023+0423 System}

Originally detected in the cluster survey of Gunn, Hoessel \& Oke (1986),
the Cl0023 system consists of four galaxy groups separated by roughly 3000
km s$^{-1}$ in radial velocity (Lubin et al.~2009a).  Two of the constituent
groups have measured velocity dispersions of 428 and 497 km s$^{-1}$, while
the second, poorer pair, have dispersions of 206 and 293 km s$^{-1}$ (Lubin et
al.~2009a).  $N$-body simulations suggest that the groups are likely bound and in
the process of forming a massive cluster within the next $\sim1$ Gyr, which,
based on virial mass estimates of the individual groups, will have a final
mass of $\sim5\times10^{14}$ M$_{\odot}$ (Lubin, Postman \& Oke 1998). 

\begin{figure}
\epsscale{1.125}
\plotone{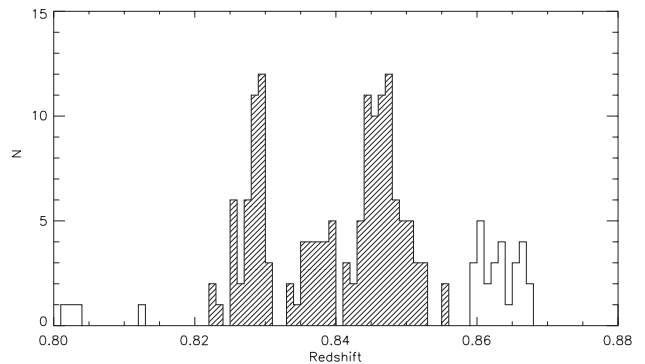}
\caption{Redshift distribution in the Cl0023 field.  Shaded regions
  correspond to galaxies within the Cl0023 structure.\label{fig-zhist}}
\end{figure}

Details of our optical observations of the Cl0023 field are presented in
Lubin et al.~(2009a).  In short, the system was imaged with the Sloan Digital
Sky Survey (SDSS) $r'i'z'$ filters using the Large Format Camera (LFC; Simcoe
et al.~2000) on the Palomar 5-m telescope and follow-up spectroscopy carried
out with the Deep Imaging Multi-object Spectrograph (DEIMOS; Faber et
al.~2003) on the Keck 10-m telescopes.  Our spectroscopic observations
yielded 423 extragalactic redshifts in the Cl0023 field and an additional 73
galaxy redshifts were incorporated from the spectroscopic survey of Oke,
Postman \& Lubin (1998). The combined catalog contains redshifts for 134
galaxies in the Cl0023 structure with $0.820<z<0.856$.   An adaptively smoothed
density map of color-selected red galaxies in the Cl0023 field constructed
from our ground-based optical imaging of the system is shown in Figure
\ref{fig-densmap}, while the redshift distribution of the system is shown in
Figure \ref{fig-zhist}.  Of the three density peaks visible in Figure
\ref{fig-densmap}, concentrations A and C consist of galaxies belonging to the
redshift peaks at $z=0.839$ and $z=0.845$, respectively, while concentration
B is a superposition of two groups along the line of sight at $z=0.828$ and
$z=0.845$.  The galaxies within the highest redshift peak at $z\sim0.864$ are
not centrally concentrated but rather extend across the entire region,
suggesting a sheet of galaxies in the near background.

\section{X-ray Observations} 
\label{sect-data}

Observations of the Cl0023 group system were carried out with {\it Chandra's}
Advanced CCD Imaging Spectrometer (ACIS; Garmire et al.~2003) on 2007 August
30 (obsID 7914).  The observation consists of a single 49.4 ks pointing of
the $16\farcm9\times16\farcm9$ ACIS-I array, with the aimpoint located at
$\alpha_{2000} = 00^{\rm h}23^{\rm m}50.9^{\rm s}$, $\delta_{2000} =
+04^{\circ}22^{\prime}55^{\prime\prime}$.  The dataset was reprocessed and
analyzed using standard CIAO 3.3 software tools\footnote{Available through
  the {\it Chandra} X-ray Center at http://cxc.harvard.edu/} in an identical
manner to the procedure detailed in Kocevski et al.~(2009a).  Images for use
in object detection were created from the level 2 event lists with a 0.492
$^{\prime\prime}$ pixel$^{-1}$ binning and corresponding spectrally-weighted
exposure maps were constructed to account for vignetting.  An adaptively
smoothed, exposure corrected image of the Cl0023 field in the soft band is
shown in Figure \ref{fig-densmap}.

\begin{figure*}
\epsscale{1.125}
\plotone{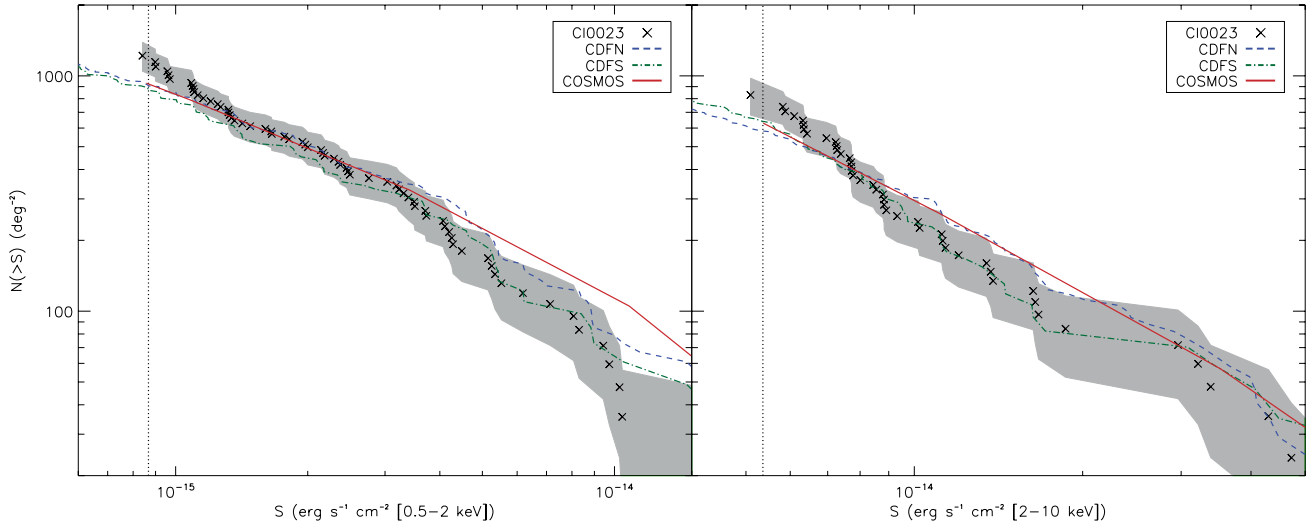}
\caption{Combined cumulative source number counts versus flux for the
  Cl0023 group merger system in the soft (0.5-2 keV;
  \emph{left}) and hard band (2-10 keV; \emph{right}).  The shaded region denotes a
  $1\sigma$ variance in the number counts.  Only sources detected above the
  $3\sigma$ level are included.  The results of an \emph{XMM-Newton} survey of the COSMOS
  field (Cappelluti et al.~2007) and that of a 130 ks \emph{Chandra} observation
  of the CDFN and CDFS are shown for comparison (solid, dashed and
  dashed-dotted lines, respectively). The vertical dotted line represents the flux at which our
  sky coverage dropped to 20\% of the full ACIS-I field of view. \label{fig-lognlogs}}
\end{figure*}

We searched for point sources using the wavelet-based {\tt wavdetect}
procedure in CIAO, employing the standard $\sqrt2$$^i$ series of wavelet
pixel scales, with $i=0-16$.   We adopted a minimum exposure threshold of
20\% relative to the exposure at the aimpoint of the observation and a
threshold significance for spurious detections of $10^{-6}$.  Object
detection was carried out on the unvignetting-corrected images and source
properties, including count rates and detection significances, were
determined with follow-up aperture photometry on the vignetting-corrected
images.  A total of 151 sources were found by {\tt  wavdetect}, of which 91
had detection significances greater than $3\sigma$ in at least one of the
0.5-2 keV (soft), 2-8 keV (hard) and 0.5-8 keV (full) bands. 

Source fluxes in the soft and hard bands were determined by normalizing a
power-law spectral model to the net count rate measured for each source.
We assumed a photon index of $\gamma = 1.4$ for the power-law
model\footnote{As this is the slope of the X-ray background (Tozzi et
  al.~2001; Kushino et al.~2002)} and a
Galactic neutral hydrogen column density of $2.66\times10^{20}$ cm$^{-2}$
(Dickey \& Lockman 1990). Full-band fluxes were determined by summing the
flux in the soft and hard bands.   Finally, rest frame X-ray luminosities
were calculated for sources matched to galaxies with measured redshifts (see
\S\ref{sect-opt-matching}) using the luminosity distance equation and a
$(1+z)^{\gamma - 2}$ $k$-correction appropriate for a power-law spectrum:

\begin{equation}
   L_{\rm X} = 4\pi d^{2}_{L} f_{\rm X} (1+z)^{\gamma - 2},
\end{equation}

\noindent where $d_{L}$ is the luminosity distance and $f_{\rm X}$ is the 0.5-8 keV X-ray
flux.


\section{Point Source Number Counts}

To calculate the cumulative number counts, $N(>S)$, of X-ray sources in
the field of the Cl0023 system, we employed the method described by Gioia et
al.~(1990):

\begin{equation}
   N(>S) = \sum_{i=1}^{N} \frac{1}{\Omega_{i}} {\rm deg}^{-2}.
\end{equation}

\noindent Here $N$ is the total number of detected point sources and
$\Omega_{i}$ is the sky area in square degrees sampled by the detector down
to the flux of the $i$th source.  The variance of the number counts was in turn calculated as

\begin{equation}
   \sigma_{i}^{2} = \sum_{i=1}^{N} \left(\frac{1}{\Omega_{i}}\right)^{2}.
\end{equation}

In order to determine $\Omega_{i}$ we constructed a flux limit map using the
method employed by Kocevski et al.~(2009a). First, all point sources detected
by {\tt wavdetect} were replaced with an estimate of the local background
with the CIAO tool {\tt dmfilth} and the resulting images binned to a pixel
scale of $32^{\prime\prime}$ pixel$^{-1}$ to produce a coarse background map.  This
map is used to determine the flux limit, $S_{\rm lim}$, for a $3\sigma$ point
source detection in any one pixel.  We can then calculate $\Omega_{i}$ by
summing the sky area covered by all pixels with $S_{\rm lim}$ equal to or
greater than the flux of the $i$th source.   

The resulting cumulative source number counts for the Cl0023 group system in
the 0.5-2 keV (left panel) and 2-10 keV (right panel) bands are shown in
Figure \ref{fig-lognlogs}.  The latter was chosen to ease comparison with
previous studies and obtained by extrapolating our 2-8 keV fluxes to 10 keV.
Also shown are the cumulative number counts measured in the COSMOS field
(Scoville et al.~2007) and the {\it Chandra} Deep Field South and North (CDFS
and CDFN; Rosati et al.~2002; Brandt et al.~2001).   The COSMOS results are
those of Cappelluti et al.~(2007) converted to a spectral index of $\gamma = 1.4$, while the CDFS and CDFN counts are the
results of our own reanalysis of single ACIS-I pointings in each
field\footnote{Observation ID numbers 581 and 2232.}.

In both the soft and hard bands, we find that the number of sources detected
in the Cl0023 field is statistically consistent ($<1\sigma$ deviation) with
the source counts observed in the reference blank fields.  In the hard band,
this agreement extends over roughly the entire sampled flux range, while in
the soft band, we find an underdensity of bright sources relative to the
COSMOS field, in agreement with the underdensity of soft sources previously
reported at these fluxes in the CDFS (Yang et al.~2003).  Our measured source
counts suggest that there is no increased nuclear activity in the Cl0023 system
detectable at X-ray wavelengths above our $3\sigma$ flux limit, which amounts
to $2.9\times10^{-15}$ erg s$^{-1}$ cm$^{-2}$ (0.5-8 keV).  At the median
redshift of the Cl0023 system, this corresponds to a rest-frame 0.5-8 keV
luminosity of $6.9\times10^{42}$ $h_{70}^{-2}$ erg s$^{-1}$, making our
observations sensitive to moderate luminosity Seyferts and QSOs in the
complex. 

To parameterize the number counts, we fit the unbinned soft- and hard-band
counts in the Cl0023 field with a power-law model of the form $N(>S) =
k(S/S_{0})^{-\alpha}$ using the maximum likelihood method of Murdoch et
al.~(1973).  Our best-fit slopes, $\alpha$, to the faint- and bright-end
number counts in the soft band are $\alpha_{\rm faint} = 0.57\pm0.12$ and
$\alpha_{\rm bright} = 1.61\pm0.30$, with a break in the distribution at
roughly $S\simeq3\times10^{-15}$ erg s$^{-1}$ cm$^{-2}$.  In the hard band we
only fit to the bright-end counts as we do not sample the faint-end
population sufficiently to obtain a separate fit.  Our best-fit slope above
$S\simeq8\times10^{-15}$ erg s$^{-1}$ cm$^{-2}$ is $\alpha_{\rm bright} =
1.69\pm0.31$.  These slopes are in good agreement (within the errors) with
previous studies of the \emph{Chandra} Deep Fields, which found Euclidean
slopes at the bright-end and $\alpha_{\rm faint} = 0.63\pm0.13$,
$0.67\pm0.14$ in the soft band for the CDFS and CDFN, respectively (Brandt et
al.~2001; Rosati et al.~2002).

\section{Optical Source Matching }
\label{sect-opt-matching}

\begin{center}
\tabletypesize{\scriptsize}
\begin{deluxetable}{cccrrr}
\tablewidth{0pt}
\tablecaption{Properties of X-ray Detected Galaxies in the Cl0023 Field with
  Measured Redshifts\label{tab-prop}}
\tablecolumns{6}
\tablehead{\colhead{RA} & \colhead{Dec} &
           \colhead{} &  \colhead{Net} &  \colhead{$F_{\rm x}$$^{\dagger}$} & \colhead{$L_{\rm x}$$^{\ddagger}$}  \\  
           \colhead{(J2000)} &  \colhead{(J2000)} &  
           \colhead{$z$} & \colhead{Counts} & \colhead{($\times10^{-15}$)} & \colhead{($\times10^{42}$)}} 
\startdata
00:23:47.5   &  04:21:17.6  &  1.487  &  7.7  &  1.49  & 12.02  \\   
00:23:43.3   &  04:18:06.3  &  0.169  &  6.4  &  2.94  &  0.21  \\    
00:23:56.3   &  04:17:60.0  &  0.683  &  3.8  &  1.14  &  1.71  \\   
00:23:55.2   &  04:25:20.2  &  1.091  & 13.5  &  4.77  & 19.86  \\   
00:24:02.5   &  04:22:12.9  &  0.442  & 39.5  & 11.29  &  6.46  \\    
00:23:51.1   &  04:27:19.8  &  0.113  & 19.7  &  5.34  &  0.17  \\   
00:23:52.2   &  04:25:53.7  &  0.682  &  5.7  &  0.69  &  1.03  \\   
00:23:48.9   &  04:21:23.7  &  0.745  & 43.0  &  9.53  & 17.23  \\   
00:23:58.5   &  04:24:51.1  &  1.336  & 10.5  &  3.76  & 24.13  \\   
\vspace*{-0.075in}
\enddata
\tablecomments{All X-ray properties measured in the 0.5-8 keV band; $^{\dagger}$ In units of erg
  s$^{-1}$ cm$^{-2}$;  $^{\ddagger}$ In units of $h_{70}^{-2}$ erg s$^{-1}$.} 
\end{deluxetable}
\end{center}
\vspace{-0.2in}

Despite the absence of a clear overdensity of X-ray sources in the Cl0023
field, we searched for AGN within the Cl0023 structure by matching our X-ray
source list to our preexisting spectroscopic catalog.  To perform this cross-correlation,
we determined the positional uncertainty associated with each X-ray source
using the empirical relationship of Kim et al.~(2007), who find that
centroiding errors increase exponentially with off-axis angle from the
aimpoint of the observation and decrease as the source counts increase with a
power-law form.  To determine the reliability of a given match, we employed a
maximum likelihood technique described by Sutherland \& Saunders (1992) and
more recently implemented by Kocevski et al.~(2009a).  The method gauges the
likelihood that a given optical object is matched to an X-ray source by
comparing the probability of finding a genuine counterpart with the
positional offset and magnitude of the optical candidate relative to that of
finding a similar object by chance.  We refer the reader to Kocevski et
al.~(2009a) for details.  

Using this technique we have matched a total of nine X-ray sources to
galaxies with measured redshifts in our spectroscopic catalog.  These
galaxies cover a broad range in redshift ($0.442 < z < 1.487$) and there is no
evidence for a concentration near the redshift of the Cl0023 system.  In fact,
we find no X-ray point sources matched to the 134 galaxies spectroscopically
associated with the four groups in the Cl0023 structure.  
However, we should note that we did not specifically target X-ray
sources with our spectroscopic observations, but instead simply cross-correlated
their positions with out existing spectroscopic database.  In future DEIMOS
observations of the system we plan to have dedicated masks for X-ray and
radio detected AGN in order to both increase our spectroscopic completeness
of X-ray sources and to determine if the lack of AGN currently observed in the Cl0023 groups holds.
The coordinates, redshifts, and X-ray properties of the nine galaxies
currently matched to X-ray point sources in the Cl0023 field are listed in Table 1.  


\section{Discussion and Conclusions}

Using \emph{Chandra} imaging of the Cl0023 complex, we have searched for
evidence of triggered nuclear activity within a dynamically active system of
four galaxy groups in the early stages of cluster formation.   Both the
redshift distribution and cumulative number counts of X-ray point sources in
the Cl0023 field reveal little evidence to suggest that the system contains X-ray
luminous AGN in excess to what is observed in the field population.  
These results are at odds with previous reports of source excesses on the outskirts
of dynamically unrelaxed clusters at high redshift.  They also appear to challange 
the notion that AGN-driven outflows play a significant role in the preprocessing 
observed in galaxy groups and environments of moderate overdensity relative to the field.
If preprocessing is underway in the Cl0023 system, our
observations suggest that powerful (quasar mode) nuclear activity is not the
predominant mechanism quenching star formation and driving the evolution of
Cl0023 galaxies.   Of course we cannot rule out a population of
low-luminosity AGN powering ``radio mode'' feedback (Croton et al.~2006) in
the Cl0023 complex as our observations are only sensitive to moderate
luminosity Seyferts and QSOs.  We are currently analyzing \emph{Very Large Array} (\emph{VLA}) 20-cm
observations of Cl0023 to search for such a population and expect to present
a full radio study of the system in a forthcoming paper (L.~M.~Lubin et al.~2009b,
in preparation). 

Our current findings are in stark contrast to the overdensity of AGN recently detected in
similar \emph{Chandra} observations of the Cl1604 supercluster at $z=0.9$,
where we find a population of Seyferts associated with an unrelaxed cluster and two rich groups 
(Kocevski et al.~2009a, 2009b).  However, the galaxy populations of these groups
differ in significant ways from those of the Cl0023 system.  The Cl1604
groups tend to have higher velocity dispersions and more evolved galaxy
populations than the Cl0023 groups, as indicated by their average SFRs and
morphological fractions (Gal et al.~2008; Lubin et al.~2009a).  Previous
observations of Cl0023 galaxies found them to be predominately late-type
systems (75\%; Lubin et al.~1998) with substantial amounts of ongoing star
formation\footnote{This is consistent with the galaxy properties of high
  redshift groups with similar velocity dispersions (e.g.~Poggianti et
  al.~2006)} (Postman, Lubin, Oke 1998; Lubin et al.~2009a), whereas the hosts
of the Cl1604 AGN tend be bulge-dominated, post-starburst galaxies which show
signs of recent or ongoing galaxy interactions.   Therefore, while Cl0023
contains galaxies which have the gas necessary to fuel nuclear activity, it apparently
lacks the bulge-dominated and massive early-type hosts in which powerful AGN
have been shown to reside (Kauffmann et al.~2003). 

A likely explanation for the absence of luminous AGN in the Cl0023 groups is
that the system lacks galaxies with sufficiently massive nuclear black holes
required to power such activity.  It has previously been shown that the
bulge-dominated S0 population in clusters and groups builds up over time at
the expense of the spiral population and that this morphological evolution is
more pronounced in lower mass systems (Poggianti et al.~2009).  There is also
evidence that these galaxies are typically more massive than their suspected
progenitors (Dressler et al.~2009), suggesting they experience growth in
their stellar bulges while in overdense environments, possibly via a
centrally concentrated burst of star formation (Dressler et al.~1999). 
Given the correlation between bulge mass and central black hole mass
(Gebhardt et al.~2000), we would expect similar growth in galactic nuclei
over the same period.  Therefore, if disruptive AGN-driven
outflows play a role in quenching star formation in groups, as has been
suggested, it may only become an important factor in the preprocessing of
galaxy populations during a later stage in the evolution of such groups and
structures, when sufficiently massive galaxies (and nuclear black holes) have
built up, but prior to hydrodynamical processes within clusters stripping
them of their gas reservoirs.   

Further observations of a larger sample of systems in the early
stages of cluster formation, with a variety of velocity dispersions and
morphological fractions, will be required to test this scenario. 
In the mean time, we are planning additional spectroscopic follow-up of the
Cl0023 groups targeting the radio bright population as well as the
remaining X-ray point sources that currently lack redshifts.  
This will give us a greater spectroscopic completeness of X-ray luminous AGN in
the Cl0023 field, which will enable us to test our current findings and should allow us to better discern
the prevalence of powerful nuclear activity during cluster formation.


\acknowledgments
This work is supported by the Chandra General Observing Program under award
number G07-8126X.  The spectrographic data used herein were obtained at the
W.M. Keck Observatory, which is operated as a scientific partnership among
the California Institute of Technology, the University of California and the
National Aeronautics and Space Administration. The Observatory was made
possible by the generous financial support of the W.M. Keck Foundation.

\bibliographystyle{apj}

\clearpage

%
%

\end{document}